\documentclass{PoS}
\pdfoutput=1

\title{Current Status of the T2K experiment}

\ShortTitle{Current Status of the T2K experiment}
       
\newcommand{\nue}                {$\nu_{e}$\xspace}

\def\nue{\nu_{e}}
\def\num{\nu_{\mu}}
\def\nut{\nu_{\tau}}

\def\lsim{\lower.7ex\hbox{${\buildrel < \over \sim}$}}
\def\gsim{\lower.7ex\hbox{${\buildrel > \over \sim}$}}

\author{\speaker{Y. Oyama}%
         \thanks{on behalf of the T2K collaboration}\\
        KEK, Japan\\
        E-mail: \email{yuichi.oyama@kek.jp}}

\abstract{
The T2K long-baseline neutrino-oscillation
experiment accumulated approximately $11.0\times 10^{20}$ POT 
(protons on target) data until June 2015.
The results of $\nu_{e}$ appearance as well as $\nu_{\mu}$
disappearance for $6.57\times 10^{20}$ POT neutrino beam data
are reported. A comparison with the results of reactor $\nu_{e}$ disappearance
experiments provides possible hints toward negative
$\delta_{\rm CP}$ phase with normal mass hierarchy. 
Preliminary results based on $4.04\times 10^{20}$ 
POT anti-neutrino beam data
are also presented.
}

\FullConference{18th International Conference From the Planck Scale to the Electroweak Scale \\
25-29 May 2015\\
Ioannina, Greece }
\begin{document}
%
%
\section{Introduction}
In the present three-flavor picture of neutrinos, the flavor eigenstates
is a mixture of mass eigenstates;

$$
\left( \begin{array}{c}\nue \\ \num \\ \nut \end{array} \right) = {\rm{\bf U_{MNS}}}
\left( \begin{array}{c}\nu_{1} \\ \nu_{2} \\ \nu_{3} \end{array} \right).
$$

\noindent
The 3$\times$3 unitary mixing matrix, ${\rm {\bf U_{MNS}}}$, is known as the 
Maki-Nakagawa-Sataka (MNS) mass matrix\cite{MNS}. It has six independent parameters:
two square-mass differences ($\Delta m^{2}_{21}$ and $\Delta m^{2}_{32}$), three mixing angles
($\theta_{12}$, $\theta_{23}$ and $\theta_{13}$) and one CP-violating phase ($\delta_{\rm CP}$).
The MNS matrix can be written as

$$
\rm {\bf U_{MNS}} = 
\left( \begin{array}{ccc}  1 & 0 & 0\\
0 & {\rm c_{23}} & {\rm s_{23}} \\
0 & -{\rm s_{23}} &  {\rm c_{23}} \end{array}\right)
\left( \begin{array}{ccc}  {\rm c_{13}} & 0 &  {\rm s_{13}}e^{-i\delta_{\rm CP}}\\
 0 & 1 & 0 \\
-{\rm s_{13}}e^{i\delta_{\rm CP}} & 0 &  {\rm c_{13}} \end{array}\right)
\left( \begin{array}{ccc}  {\rm c_{12}} & {\rm s_{12}} & 0\\
{\rm -s_{12}} & {\rm c_{12}} & 0  \\
0 & 0 &  1 \end{array}\right)
$$
where $\rm{c}_{ij} = \cos\theta_{ij}$ and $\rm{s}_{ij} = \sin\theta_{ij}$. The ranges of the oscillation parameters have been reported in many experiments, and
are summarized by the particle data group\cite{PDG2014} as listed in Table~\ref{tab:oscparam}. 

\begin{table}[b!]
\caption{
Neutrino oscillation parameters reported in many experiments, and
summarized by the particle data group\cite{PDG2014}.
NH indicates normal mass hierarchy, $m_{3} > m_{2} > m_{1}$, and IH indicates
inverted mass hierarchy,  $m_{2} > m_{1} > m_{3}$.
}
\smallskip
\begin{center}
\begin{tabular}{ll}
\hline
\hline
$\sin^{2}2\theta_{12}$  &  $~~~=~~~~~0.846\pm 0.021$\\ 
$\Delta m^{2}_{21}$      &  $~~~=~~~~~(7.53\pm 0.18)\times 10^{-5}{\rm eV^{2}}$\\
$\sin^{2}2\theta_{23}$ & $~~~=~~~~~0.999^{+0.001}_{-0.018}{\rm(NH)}$\\
$\sin^{2}2\theta_{23}$ & $~~~=~~~~~1.000^{+0.000}_{-0.017}{\rm(IH)}$\\
$\Delta m^{2}_{32}$ & $~~~=~~~~~(2.44\pm 0.06)\times 10^{-3} {\rm eV^{2}}{\rm (NH)}$\\
$\Delta m^{2}_{23}$ & $~~~=~~~~~(2.52\pm 0.07)\times 10^{-3} {\rm eV^{2}}{\rm (IH)}$\\
$\sin^{2}2\theta_{13}$ & $~~~=~~~~~(0.093\pm  0.008)$\\
\hline
\hline
\end{tabular}
\end{center}
\label{tab:oscparam}
\end{table}

Among these measurements, the first indication of non-zero $\theta_{13}$
was claimed by the T2K (Tokai to Kamioka) experiment in 2011,
based on $1.43\times 10^{20}$ protons on target (POT)\cite{T2K2011}.
The result was confirmed by another long-baseline neutrino beam experiment\cite{MINOS},
and three reactor experiments\cite{reactors}.

After the measurements of non-zero $\theta_{13}$, there are still open
questions in neutrino oscillation physics. The two most important questions involve determining
whether CP violation exists in the neutrino sector (the value of the CP phase) and defining the neutrino
mass hierarchy,  in other words, the sign of $\Delta m^{2}_{32}$.

\bigskip

T2K\cite{T2KLOI} is a long-baseline neutrino-oscillation experiment that began in 2009.
A high-intensity neutrino beam from the J-PARC Main Ring is directed toward
the Super-Kamiokande (SK) detector, 295~km away.
This article provides updated results from the T2K experiment based on data accumulated until June 2015.

\section{T2K neutrino beam line and detectors}

\begin{center}
\begin{figure}[h]
\hspace{2pc}
\includegraphics[width=35pc]{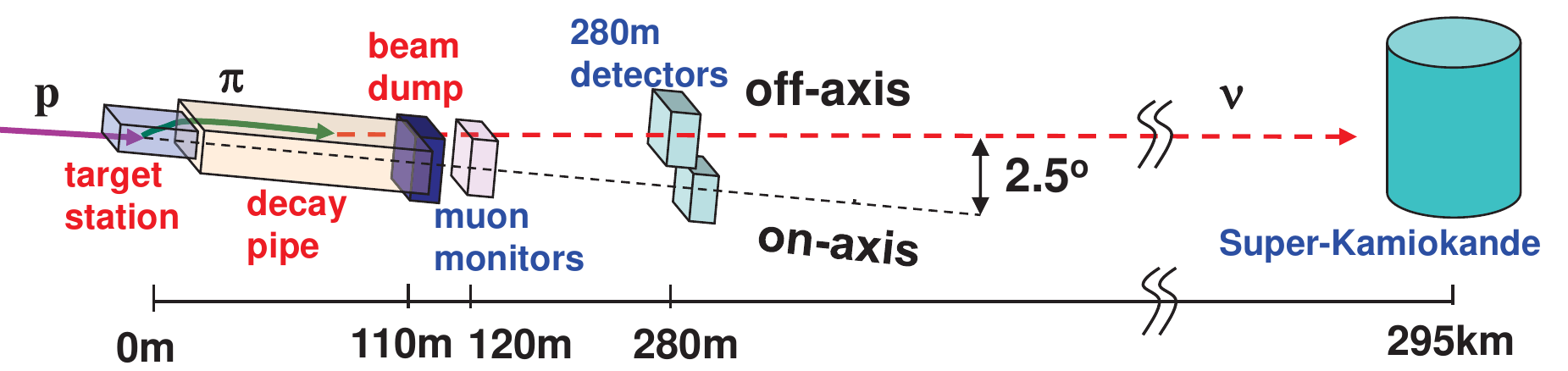}
\caption{\label{fig:concept}A schematical view of the T2K neutrino beamline and detectors. Beamline components are shown
in red letters, and detectors are shown in blue letters.}
\end{figure}
\end{center}

A schematical view of the T2K neutrino beamline and detector components is shown in Fig.\ref{fig:concept}.
Details about them were previously reported in \cite{t2knim}.
This article presents some of the important features.

\subsection{Neutrino beam line}
The proton beam for the T2K experiment is extracted from the J-PARC 30GeV Main Ring proton synchrotron.
It is delivered to the carbon target in the target station.
Pions and kaons are produced and subsequently bent in the direction of SK by a magnetic field
produced by magnetic horns.
Neutrinos are produced as decay products of pions and kaons.
All particles except neutrinos and high energy ($>$ 5GeV) muons are absorbed by a beam dump
located 110m downstream of the target.

The most important feature of the T2K neutrino beam line is the off-axis beam,
which was originally proposed in \cite{offaxis}.
The conceptual idea regarding the off-axis angle in T2K
was precisely reported in \cite{crimea}.
When the beamline construction was started,
the off-axis angle was tunable\cite{crimea} because
the best off-axis angle was unknown.
Based on information about the $\Delta m^{2}_{32}$ from other experiments\cite{SKatm,MINOS2006},
the off-axis angle was adjusted to 2.5$^{\circ}$ in 2007.
The corresponding peak energy of the neutrino flux is approximately  600MeV,
and the oscillation study is most sensitive around $\Delta m^{2}_{32} = 2.5\times 10^{-3}{\rm eV^{2}}$.

\subsection{Muon monitors}
Two types of muon monitors are installed downstream of the beam dump.
They are an Ionization Chamber and a Semiconductor Array.
They can measure the intensity distribution of muons that have escaped from the beam dump.
Because these muons are mainly produced along with neutrinos from the $\pi^{+}$ decays,
their beam center is in accordance with the beam center of neutrinos. 
The position of the beam center can be monitored on a bunch-by-bunch basis
within a 3~cm resolution from the peak of the muon intensity distribution.
This position resolution corresponds to a beam direction accuracy of 0.25~mrad.

\subsection{Near detectors}
The near detectors were constructed in an underground experiment hall of
33.5m depth and 17.5m diameter at 280m downstream from the target.
Two detectors were installed; an on-axis detector (aimed in the direction of the
neutrino beam center), and an off-axis detector (aimed in the direction of SK).
\medskip

\begin{figure}[b!]
\center{{\includegraphics[height=6.0cm]{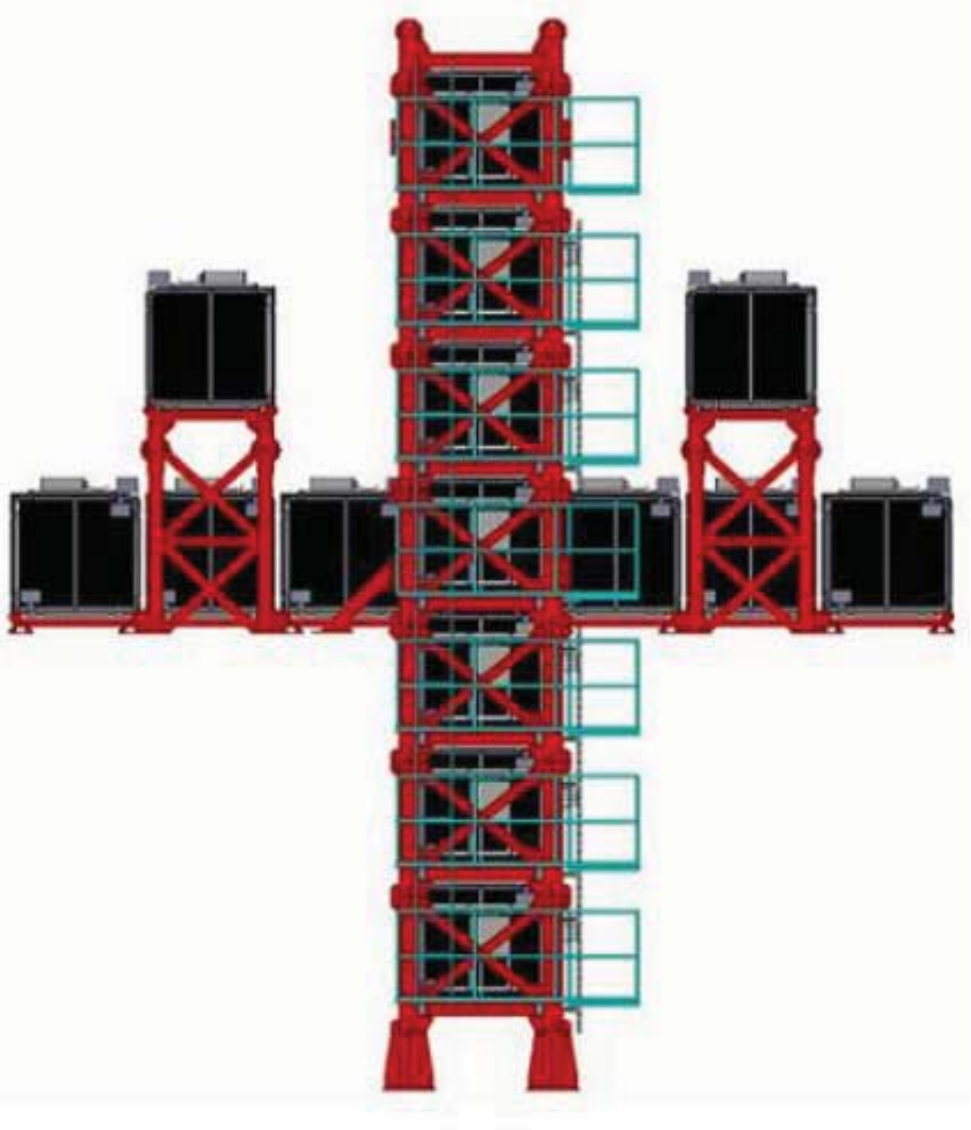}}
\hskip 2.0cm
        {{\includegraphics[height=6.0cm]{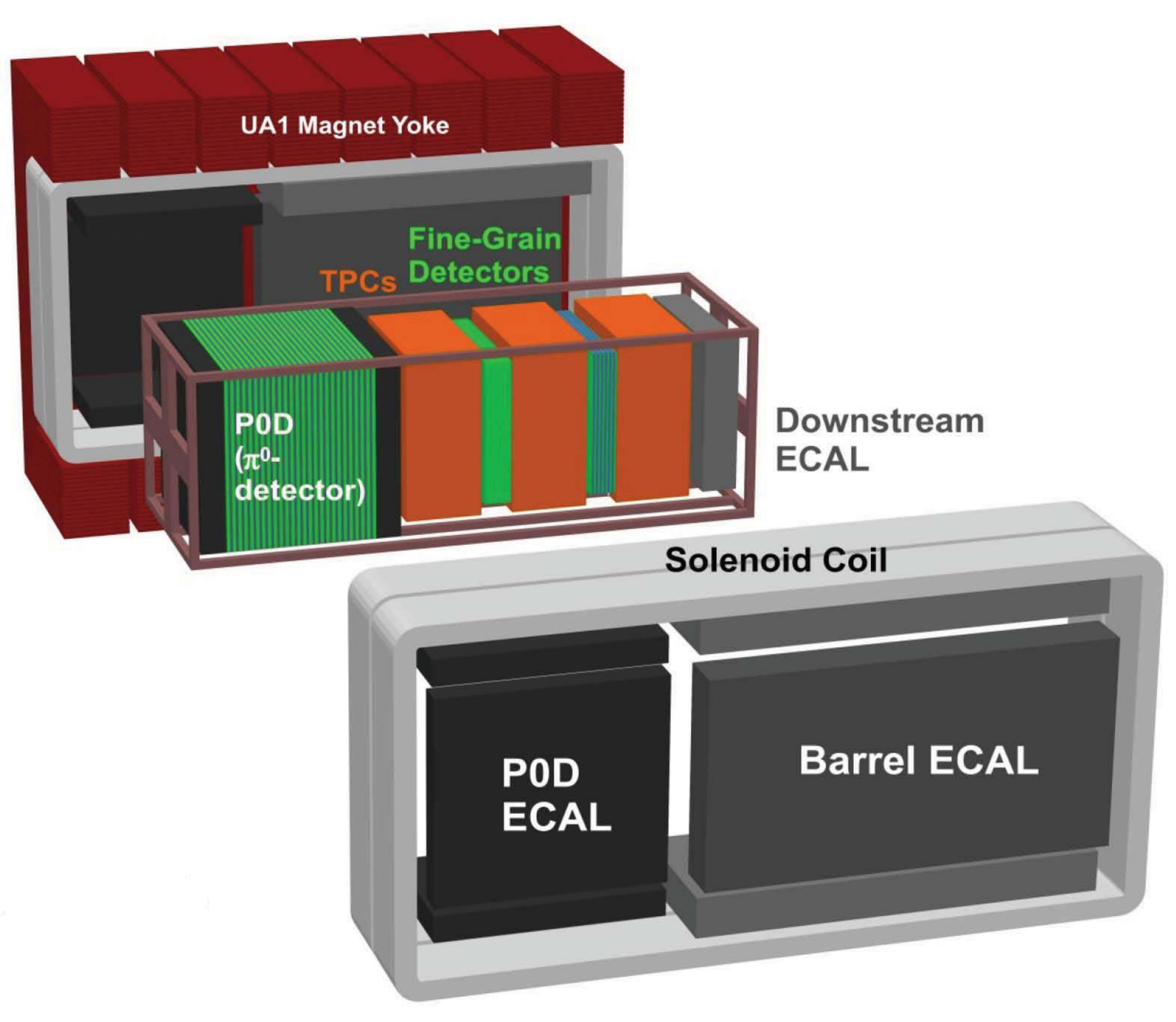}\vskip 0.5cm}}
}
\center{
\caption{\label{fig:near}
Schematic view of the INGRID on-axis detector (left), and
the ND280 off-axis detector (right).
}}
\end{figure}

The on-axis detector, namely the INGRID detector, consists of
16 1m$\times$1m$\times$1m cubic modules as shown in Fig.\ref{fig:near}(left).
Each module is a "sandwich" of
11 scintillator layers and 10 iron layers.
They are surrounded by four veto planes.
The modules are arranged as follows:
seven horizontally, seven vertically, and two off-diagonally.
The neutrino beam center can be measured based on the horizontal/vertical distribution
of the neutrino event rate.

\medskip

The off-axis detector is named ND280.
A schematical view of the ND280 detector is shown in Fig.\ref{fig:near}(right).
The ND280 detector's purposes as a near detector
are to measure the energy spectrum of muon neutrinos in SK's direction,
and to measure the fractions of electron neutrinos.

All detector components except SMRD are placed inside a 0.2~T magnetic field produced by the recycled UA1
magnet from CERN.
\noindent
The P0D (Pi-zero Detector) is a subdetector component that is placed upstream inside
the magnet. It is a "sandwich" of scintillator planes, lead plates, and a water target.
It is customized for the measurement of neutral $\pi^{0}$ production. $\gamma$-rays from
$\pi^{0}\rightarrow 2\gamma$ are converted to electromagnetic showers by lead
plates, and are detected by scintillators.

Downstream of the P0D, three TPCs (Time Projection Chambers) and
two FGDs (Fine Grained Detectors) are placed.
The TPCs can measure the momentum of muons from the curvature of the muon track in the magnetic field.
The momentum resolution is greater than 10\% at 1~GeV.
FGDs consist of scintillator bars. They provide the target material for
neutrino interactions,
and are optimized for detecting the proton recoils.
By combining TPCs and FGDs, the energy spectrum of $\nu_{\mu}$
can be precisely measured from CCQE (Charged Current Quasi-Elastic) neutrino interactions.

Other detector components, SMRD and ECAL, are installed in the area surrounding the P0D and TPC/FGD.
They are precisely reported in \cite{t2knim}.  

The ND280 detector can be used as a near detector for the long-baseline neutrino-oscillation experiment,
and as an independent detector for neutrino cross section measurements.
Some results on the cross section measurements can be found in \cite{crosssection}.  
\subsection{Super-Kamiokande}
The far detector, Super-Kamiokande (SK), is a 50kton water Cherenkov detector\cite{SKdetector}
located 1000m underground in the Kamioka mine, Japan.
Its distance from J-PARC is 295km.
In the inner detector (ID), 22.5kton of fiducial volume
are viewed by 11,129 20-inch diameter PMTs.
The outer detector (OD), which surrounds the ID, is also a water Cherenkov detector.
It is used to veto events that enter or exit the ID.
SK started its operation in April 1996. After a complete upgrade of its electronics systems in 2008,
it was named SK-IV.

The most important characteristic of SK, as the far detector 
of the T2K experiment, is its ability to differentiate between muons and electrons.
This particle identification directly implies identification
between the parent $\nu_{\mu}$ (or ${\bar \nu_{\mu}}$) and $\nu_{e}$ (or ${\bar \nu_{e}}$).
The principal of particle identification was reported in \cite{Kasuga}.
Muons passing through the detector are often unscattered due to their relatively large mass,
and thereby produce clear ring patterns. Electrons, in contrast, scatter
and produce electromagnetic showers, resulting in a diffuse
ring edge.
It was verified that probability of the $\mu$/$e$ misidentification is less than 1\%\cite{Kasuga}. 
\section{Status of the data-taking process}

\begin{center}
\begin{figure}[b!]
\hskip 1.7cm
\includegraphics[width=30pc]{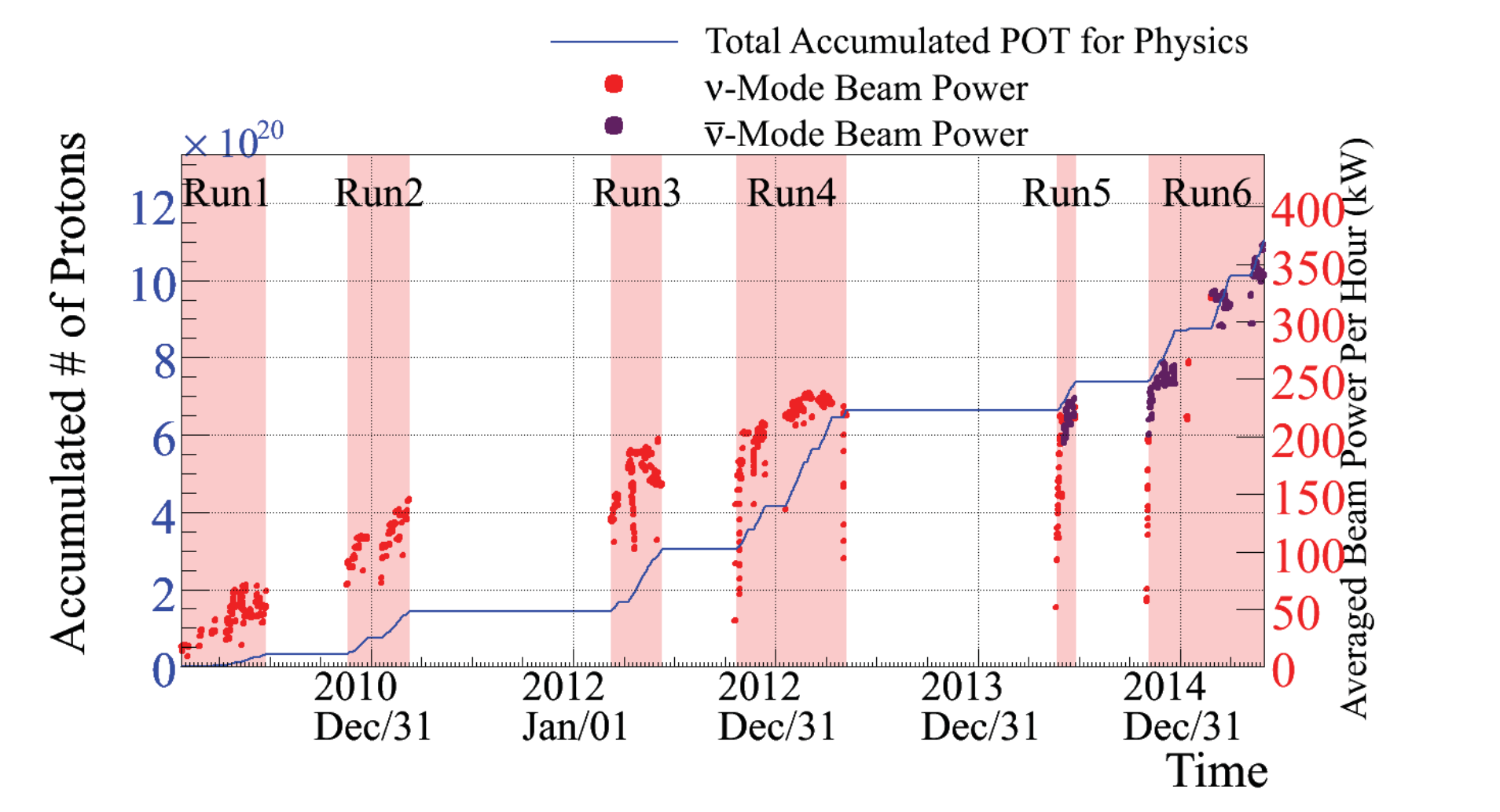}
\caption{\label{fig:beamhistory}History of primary proton beam intensity in the T2K experiment.
Red (violet) dots show averaged beam power per hour in the neutrino (anti-neutrino) mode beam;
the scale is given in the right vertical axis.
The blue solid line shows the accumulated number of delivered protons from the beginning of the experiment.
The scale is given in the left vertical axis.
}
\end{figure}
\end{center}

The history of the proton beam delivery is shown in Fig.\ref{fig:beamhistory}.
The physics data-taking began in January 2010.
In early 2010, one beam pulse had six bunches in $\sim$5$\mu$s.
The number of protons per pulse (ppp) was $\sim 2\times10^{13}$~ppp,
and the beam pulse cycle time was 3.52~seconds.

Many efforts were made by the J-PARC accelerator group to upgrade the beam power.
In June 2015, the number of bunches in each
pulse was 8, and the number of protons per pulse was $\sim 1.8\times10^{14}$~ppp.
The beam cycle time was upgraded to 2.48~seconds.
The maximum beam power achieved through June 2015 was 371~kW.

We accumulated 11.0$\times$10$^{20}$ POT data until June 4, 2015.
This is about 14\% of our goal of $\sim 78\times$10$^{20}$~POT, which can be attained over
five years of beam operation at a power of 750kW.

In June 2014, the direction of the magnetic horn current was reversed and
an anti-neutrino beam run was started.
The magnetic horns focus on the negative pions instead
of the positive pions, and anti-neutrinos (as decay products of
negative pions) can travel in a forward direction.

\begin{center}
\begin{figure}[b]
\center{
        {\includegraphics[width=36pc]{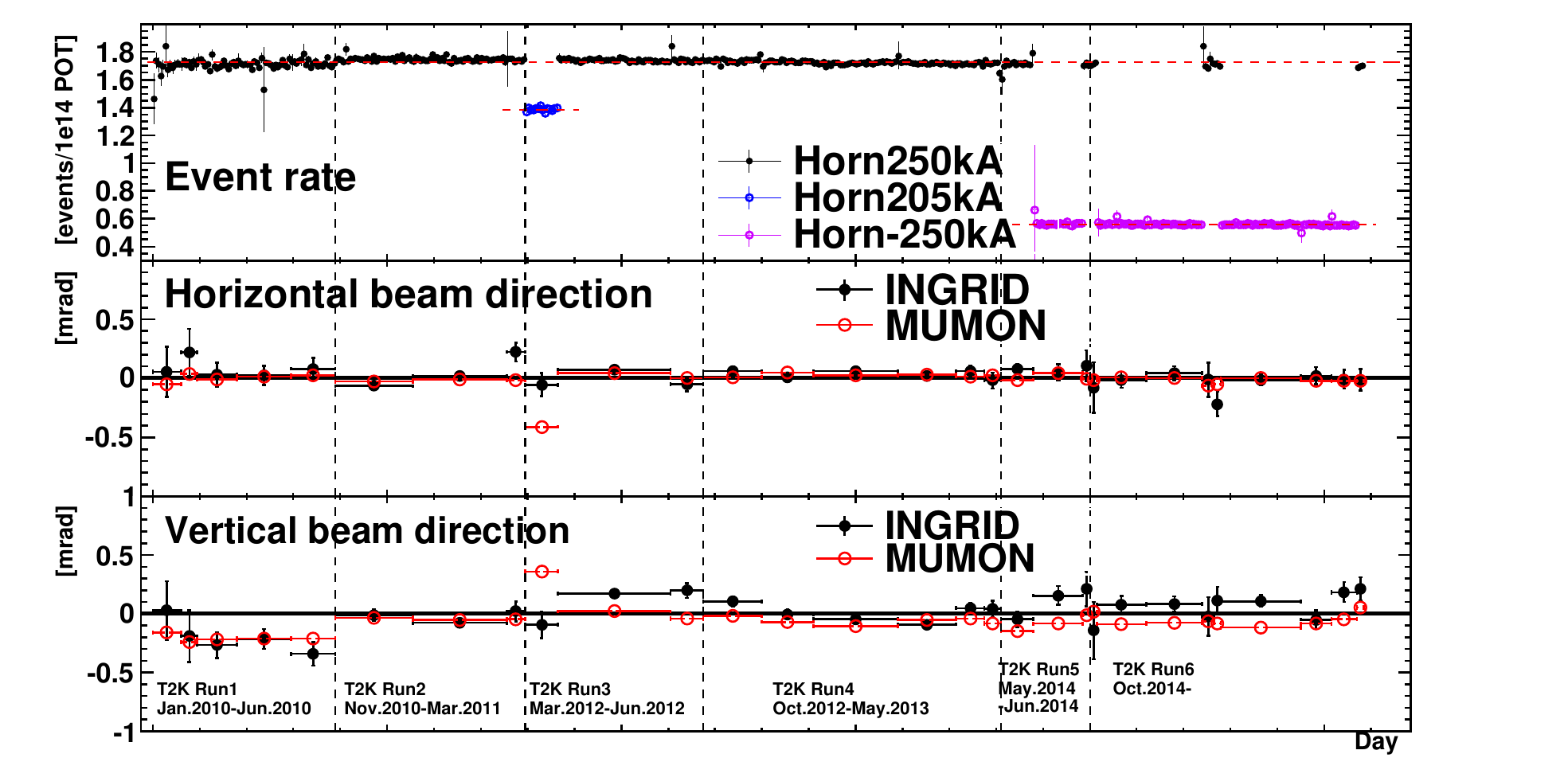}}
}
\caption{\label{fig:stability}
The top figure shows the event rate observed by the INGRID detector
normalized by POT. 
The other two figures show muon beam directions measured
by the INGRID detector and muon monitors.
The beam direction is well controlled and is stable within 1mrad.
}
\end{figure}
\end{center}

The stability of the beam direction is continuously monitored by the muon monitors
and the INGRID detector as shown in Fig.~\ref{fig:stability}.
The beam direction is stable and well controlled within $\pm$1~mrad accuracy.
Note that if the beam direction is changed by 1~mrad, the neutrino flux at SK is changed
by 2\% at $E_{\nu}=0.5\sim 0.7$GeV. Therefore, the accuracy of the beam direction
is satisfactory.
The stability of the neutrino event rate observed in the INGRID detector is also
shown in Fig.~\ref{fig:stability}.

In the next four sections, official results based on 6.57$\times10^{20}$POT neutrino beam data accumulated
until May 2013 are reported\cite{T2Kdisapp,T2Kapp,T2Kcombine}.
In sections 8$\sim$10, physics motivation and preliminary results based on 4.04$\times10^{20}$POT
anti-neutrino beam data recorded until June 2015 are reported\cite{EPS2015}.

\vskip 2.0cm
\section{Neutrino event selection}

Beam-related neutrino events
are selected from SK data recorded during  a 6.57$\times10^{20}$POT beam period.
The event selection process comprises two steps.
The first step is the same for $\nu_{\mu}$ and $\nu_{e}$, and
beam-related fully contained fiducial volume (FCFV) events are selected.
Separate conditions are applied for $\nu_{\mu}$ and $\nu_{e}$ selection
in the second steps.
Details of the event selection are provided below.

Beam-related Fully Contained (FC) events are selected according to the following three conditions.
\begin{description}
\item[~~~~] 1) Total energy deposited in the ID is greater than 30MeV
\vspace{-3mm}
\item[~~~~] 2) No outer detector activity
\vspace{-3mm}
\item[~~~~] 3) SK event time is within a range of -2 to 10$\mu$sec from the beam trigger time

\end{description}
The third condition examines correlations of the GPS time recorded in Tokai and in SK.
A spatial reconstruction algorithm is applied to the FC events, and the following condition is applied. 
\begin{description}
\item[~~~~] 4) The vertex position should be in the 22.5kton of fiducial volume
\end{description}

\noindent
Note that the above analysis procedures except for step 3 are almost the same
as the well-established atmospheric neutrino analysis\cite{SKatm}.

\begin{table}[b!]
\caption{
Conditions of $\nu_{\mu}$ and $\nu_{e}$ event selection.
Number of selected events for 6.57$\times$10$^{20}$POT T2K data and
the expectation for no oscillation are also listed.
}
\smallskip
\begin{center}
\begin{tabular}{l|l|l}
\hline
\hline
                          &~~~~~~~~~~$\nu_{\mu}$ selection    &~~~~~~~~~~$\nu_{e}$ selection     \\
\hline
Conditions & \multicolumn{2}{l}{1)Total energy deposit in the ID > 30~MeV}\\
               & \multicolumn{2}{l}{2)No outer detector activity}\\
               & \multicolumn{2}{l}{3)SK event time agrees with the beam trigger time}\\
               & \multicolumn{2}{l}{4)The vertex position should be in the 22.5~kton fiducial volume}\\
               & \multicolumn{2}{l}{}\\
               & $\mu$-1)$\mu$-like single ring & $e$-1)$e$-like single ring\\ 
               & $\mu$-2)$p_{\mu} > 200$~MeV/c & $e$-2)$p_{e} > 100$~MeV/c\\ 
               & $\mu$-3)0 or 1 delayed-electron signal& $e$-3)0 delayed-electron signal\\ 
               & & $e$-4)$E_{\nu}^{rec} < $1250~MeV\\ 
               & & $e$-5)Tight $\pi^{0}$ rejection\\ 
\hline
Number of selected events      &~~~~~~~~~~120  &~~~~~~~~~~28 \\
\hline
Number of expected events      &~~~~~~~~~~446 $\pm$23 (sys.)  &~~~~~~~~~~4.9 $\pm$0.6 (sys.) \\
\hline
\hline
\end{tabular}
\end{center}
\label{tab:reduction}
\end{table}

After these conditions, 377 events are selected as
FCFV (Fully Contained Fiducial Volume) events.
The expected number of background events from
non beam-related sources in accidental coincidence is
estimated to be 0.0085.

A ring counting algorithm and the particle identification algorithm are 
applied to the 377 FCFV events.
The events are categorized into
single ring $\mu$-like events, single ring $e$-like events and multi-ring
events.

\bigskip

In the $\nu_{\mu}$ event selection, the following three conditions are applied.

\begin{description}
\item[~~~~] $\mu$-1) $\mu$-like single ring
\vspace{-3mm}
\item[~~~~] $\mu$-2) $p_{\mu} > 200$~MeV/c
\vspace{-3mm}
\item[~~~~] $\mu$-3) 0 or 1 delayed-electron signal
\end{description}

\noindent
In $\mu$-3), the delayed-electron signal is thought to be a Michel electron.
After these conditions, 120 events remain as $\nu_{\mu}$ candidates.

\bigskip

In the $\nu_{e}$ event selection, five conditions are applied.

\begin{description}
\item[~~~~] $e$-1) $e$-like single ring
\vspace{-3mm}
\item[~~~~] $e$-2) $p_{e} > 100$~MeV/c
\vspace{-3mm}
\item[~~~~] $e$-3) 0 delayed-electron signal
\vspace{-3mm}
\item[~~~~] $e$-4) $E_{\nu}^{rec} < $1250~MeV
\vspace{-3mm}
\item[~~~~] $e$-5) Tight $\pi^{0}$ rejection
\end{description}

\noindent
By the condition $e$-3), events accompanied by Michel electrons are rejected.
In the condition $e$-4), the neutrino energy $E_{\nu}^{rec}$ is calculated by assuming quasi-elastic
kinematics. To suppress contributions by the intrinsic $\nu_{e}$ components arising
primarily from kaon decays, the event is rejected if $E_{\nu}^{rec}~>~1250$MeV.
The condition $e$-5) is a special algorithm that suppresses contamination by misidentified $\pi^{0}$.
The reconstruction of the second ring is forced,
and the 2-ring invariant mass, $M_{inv}$, is calculated.
$M_{inv}$ and the likelihood for $\pi^{0}$ are examined, and $\pi^{0}$-like events are rejected
with tight criteria.
Finally, 28 events remain after all selection criteria.

The conditions of the event selection and the results of the selection are summarized in Table~\ref{tab:reduction}.

\section{Expected events and their systematic errors}
The number of expected neutrino events in SK is carefully calculated by
a Monte Carlo simulation of neutrino productions and their interactions.

\begin{figure}[t!]
\center{
        {\includegraphics[width=36pc]{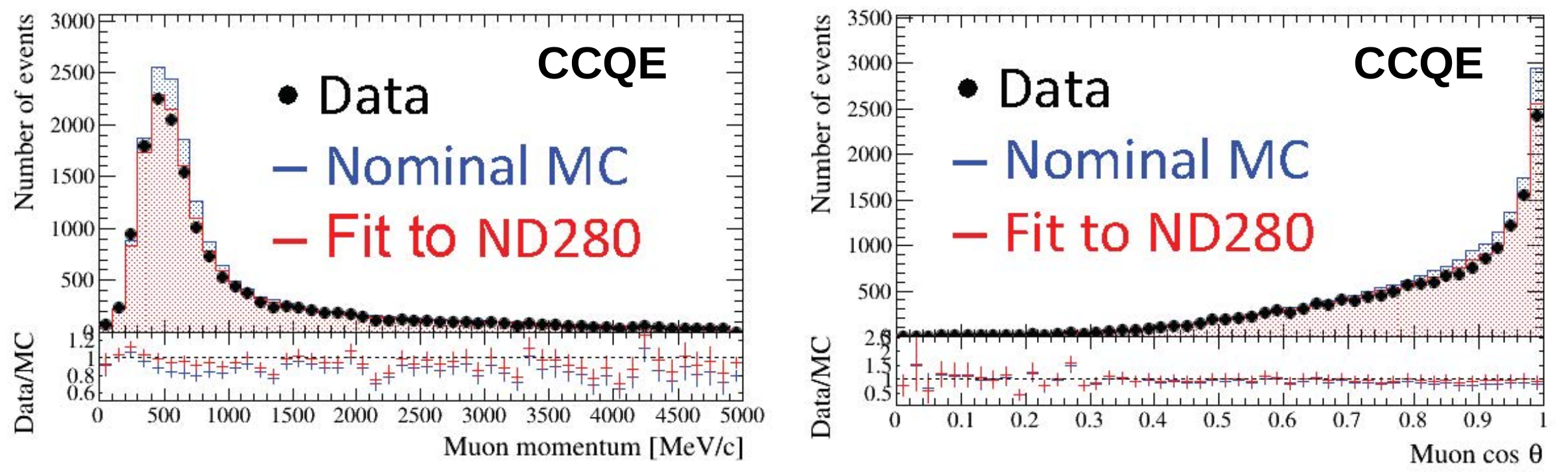}}
}
\center{
\caption{\label{fig:nd280ana}
Momentum (left) and angle (right) distribution of CCQE events in ND280.
After the adjustment of systematic uncertainties, agreements between the data and Monte Carlo simulation
are excellent.
}}
\end{figure}

From the interactions between primary proton beam and the carbon target,
secondary pions and kaons are generated.
Pions and kaons propagate in the secondary beamline.
Neutrinos are produced as decay products of pions and kaons.
All of these neutrino production processes are simulated by
standard simulation programs, FLUKA\cite{FLUKA1,FLUKA2},
GEANT3\cite{GEANT3} and GCALOR\cite{GCALOR}.
To reduce the systematic uncertainties in pion and kaon productions,
some of the T2K members joined the CERN NA61 experiment: "Study of hadron
productions in hadron-nucleus and nucleus-nucleus collisions in CERN SPS."
Pion\cite{NA61p} and kaon\cite{NA61k} production data from a 30GeV proton beam and carbon target
from NA61 are employed in the simulation.
Systematic errors on neutrino fluxes in the 0.1GeV$\sim$5GeV range at ND280 and SK
are in the 10$\sim$15\% range. 
However, fluxes at the two detectors are highly correlated,
and some of the systematic errors are common.
An extrapolation of the ND280 analysis can reduce systematic errors in SK as discussed below.

The neutrino interactions are simulated with the NEUT\cite{NEUT} neutrino interaction generator.
The GENIE\cite{GENIE} generator is also used to evaluate uncertainties caused by the simulation program.
Model parameters in the neutrino interaction simulation, such as axial vector mass, are tuned by using
external neutrino data. Results from MiniBooNe\cite{MiniBooNe} are used as primary inputs.
At present, nominal systematic errors for the neutrino cross section are larger than $\sim$10\%. 

To reduce systematic errors, ND280 data are used.
CCQE (Charged Current Quasi-Elastic) events, CC1$\pi$ (Charged Current One Pion) events,
and CCdis (Charged Current deep inelastic scattering) events are selected based
on track topologies in ND280 FGD/TPC. 
$p_{\mu}$ and $\cos\theta_{\mu}$ distributions from the data and the simulation
are carefully compared.
Based on the comparison, all systematic errors related to cross sections and neutrino fluxes are adjusted.
After the parameter adjustment\cite{T2Kdisapp,T2Kapp},
the agreements are excellent as shown in Figure~\ref{fig:nd280ana}.

The adjusted parameters in the ND280 analysis can be also applied for SK.
After the use of parameter adjustments, 
the systematic errors on expected neutrino events at SK are 
reduced from 23.5\% to 7.7\% for $\nu_{\mu}$ candidates,
and from 26.8\% to 6.8\% for $\nu_{e}$ candidates.
The extrapolation of ND280 analysis strongly reduces the systematic errors. 

The event selection algorithm presented in section 4 was also
applied to the Monte Carlo events.
Numbers of expected neutrino event candidates for no oscillation are 446$\pm$23~(sys.) for
$\nu_{\mu}$ and 4.9$\pm$0.6~(sys.) for $\nu_{e}$.
The results are shown in Table~\ref{tab:reduction}.
\section{Results of muon neutrino disappearance}

\begin{figure}[t!]
\center{
        {\includegraphics[width=36pc]{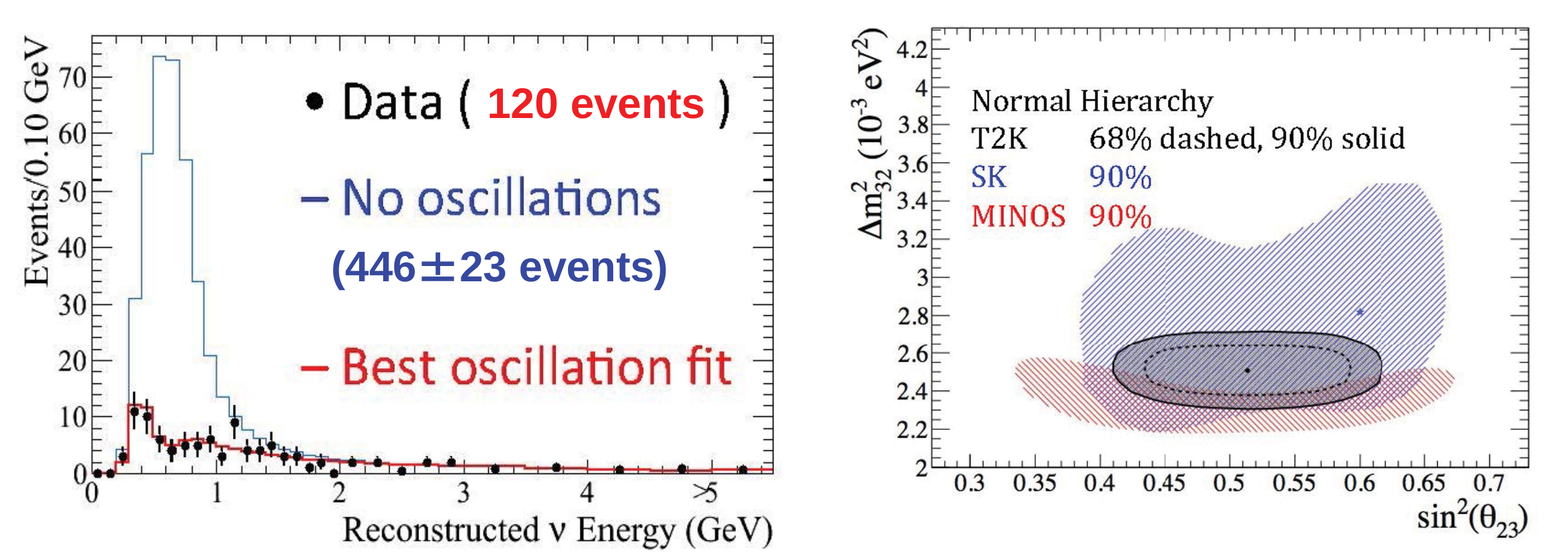}}
}
\center{
\caption{\label{fig:numu}
(left)~Distribution of reconstructed neutrino energy for 120 muon neutrino candidates observed in
$6.57\times10^{20}$POT data. The expectations for no oscillation and for best-fit oscillation parameters
are also shown.    
(right)~Constraints on oscillation parameters $\Delta m^{2}_{32}$ and $\sin^{2}\theta_{23}$ obtained from
the disappearance of muon neutrino candidates. Normal mass hierarchy is assumed.
Constraints from SK\cite{SKdisapp} and MINOS\cite{MINOSdisapp} are also shown.
}}
\end{figure}

As reported in the previous sections, 120 muon neutrino candidates
are observed in $6.57\times10^{20}$ POT data, in which 446$\pm$23 events are expected if no oscillation is assumed.
The neutrino energy distribution for 120 events is also shown in Figure~\ref{fig:numu} (left).
The disappearance of muon neutrino events as well as the distortion of the neutrino energy
spectrum are obvious.

The best oscillation parameters are calculated by comparing expected events for given
oscillation parameters and the data.
The best oscillation parameters were calculated as follows:
$$
\begin{array}{ll}
\Delta m^{2}_{32} &= (2.51\pm0.10)\times 10^{-3} {\rm eV^{2}}\\
\sin^{2}\theta_{23} &= 0.514^{+0.055}_{-0.056}\\
\end{array}
$$
\noindent
for normal mass hierarchy, and
$$
\begin{array}{ll}
\Delta m^{2}_{32} &= (2.48\pm0.10)\times 10^{-3} {\rm eV^{2}}\\
\sin^{2}\theta_{23} &= 0.511\pm 0.055\\
\end{array}
$$
\noindent
for inverted mass hierarchy.

The constraint in the two dimensional $\sin^{2}\theta_{23}$ - $\Delta m^{2}_{32}$ plane for normal mass hierarchy
is shown in Figure~\ref{fig:numu}~(right), together with results from SK\cite{SKdisapp} and MINOS\cite{MINOSdisapp}.
The T2K results are consistent with those of the two experiments, and provides the
most stringent constraints for $\sin^{2}\theta_{23}$.

\section{Results of electron neutrino appearance}

As presented in sections 4 and 5, 28 electron neutrino candidates are observed
where $4.9\pm 0.6$ events are expected for no oscillation.
From a statistical calculation,  the significance of the signal is 7.3 times the standard deviation.
It can be concluded with certainty that this is {\bf discovery} of the electron neutrino
appearance. (Note that the T2K results based on $1.43\times 10^{20}$POT data collected until 2011 was
claimed as indication\cite{T2K2011,oyama2011}, and the results based on the $3.01\times 10^{20}$POT data
collected until 2012 was claimed as evidence\cite{T2K2013}.)

\begin{center}
\begin{figure}[t!]
\vskip 1.0cm
\includegraphics[width=41pc]{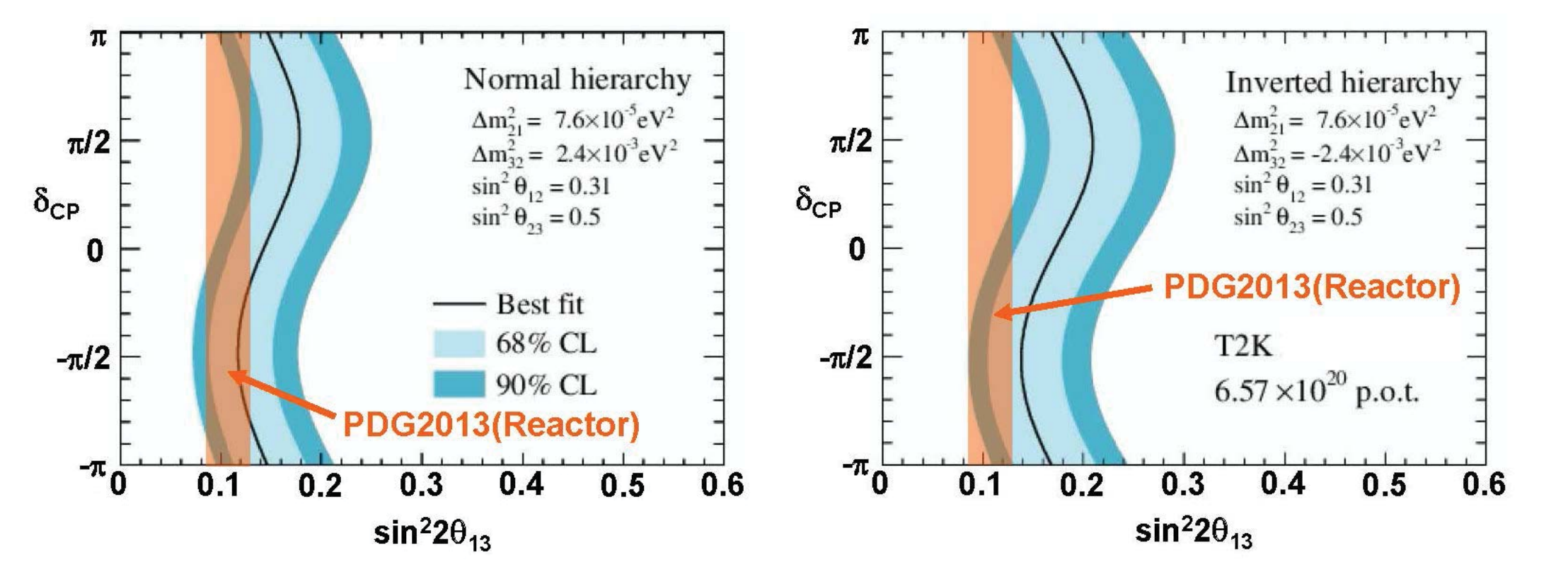}
\caption{\label{fig:nueapp}
Constraints on oscillation parameters in $\sin^{2}2\theta_{13}$ - $\delta_{\rm CP}$ plane.
Normal mass hierarchy (left) and inverted mass hierarchy (right) are assumed.
Constraints on $\sin^{2}2\theta_{13}$ from reactor experiments are also shown.
}
\end{figure}
\end{center}

Constraints on oscillation parameters are calculated by comparing the data
and expectations.
The allowed region in the $\sin^{2}2\theta_{13}$ - $\delta_{\rm CP}$ plane
for normal mass hierarchy and inverted mass hierarchy
are shown in Figure~\ref{fig:nueapp}.
The 68\% confidence level intervals for $\sin^{2}\theta_{13}$  are

$$
\begin{array}{llll}
\sin^{2}2\theta_{13} &= 0.140^{+0.038}_{-0.032}~~~~~~&{\rm for~normal~mass ~hierarchy~(NH)}&~~~~~~{\rm and}\\
&&&\\
\sin^{2}2\theta_{13} &= 0.170^{+0.045}_{-0.037}~~~~~~&{\rm for~inverted~mass~hierarchy~(IH)}& \\
\end{array}
$$
\noindent
if absence of a CP violation phase is assumed (i.e. $\delta_{\rm CP}=0$).

The constraints for $\sin^{2}2\theta_{13}$ from reactor experiments\cite{PDG2013} are also
shown in Figure~\ref{fig:nueapp}.
The overlap between T2K and the reactor results indicates that
negative $\delta_{\rm CP}$ with normal mass hierarchy is favored.
More complicated and exhaustive statistical analysis methods\cite{T2Kcombine} were employed, 
in which the muon neutrino disappearance and electron neutrino appearance are
combined, and four parameters ($\delta_{\rm CP}$, $\theta_{13}$, $\theta_{23}$
and $\Delta m^{2}_{32}$) are fitted simultaneously.
From the analysis, the parameter ranges

$$
\begin{array}{llll}
0.146\pi & < \delta_{\rm CP} < 0.825\pi & \rm{~~~~~~for~normal~mass~hierarchy~(NH)} &~~~~{\rm and}\\
&&&\\
-0.080\pi & < \delta_{\rm CP} < 1.091\pi & \rm{~~~~~~for~inverted~mass~hierarchy~(IH)} &\\
\end{array}
$$
\noindent
are excluded with a 90\% confidence level.

These results are considered to be {\bf the first hints toward
${\bf \delta_{\rm CP}}{\bf \sim}{\bf -\pi/2}$
and normal mass hierarchy}.

\vfill\eject

\section{Motivation for anti-neutrino beam}
\begin{center}
\begin{figure}[t!]
\includegraphics[width=38pc]{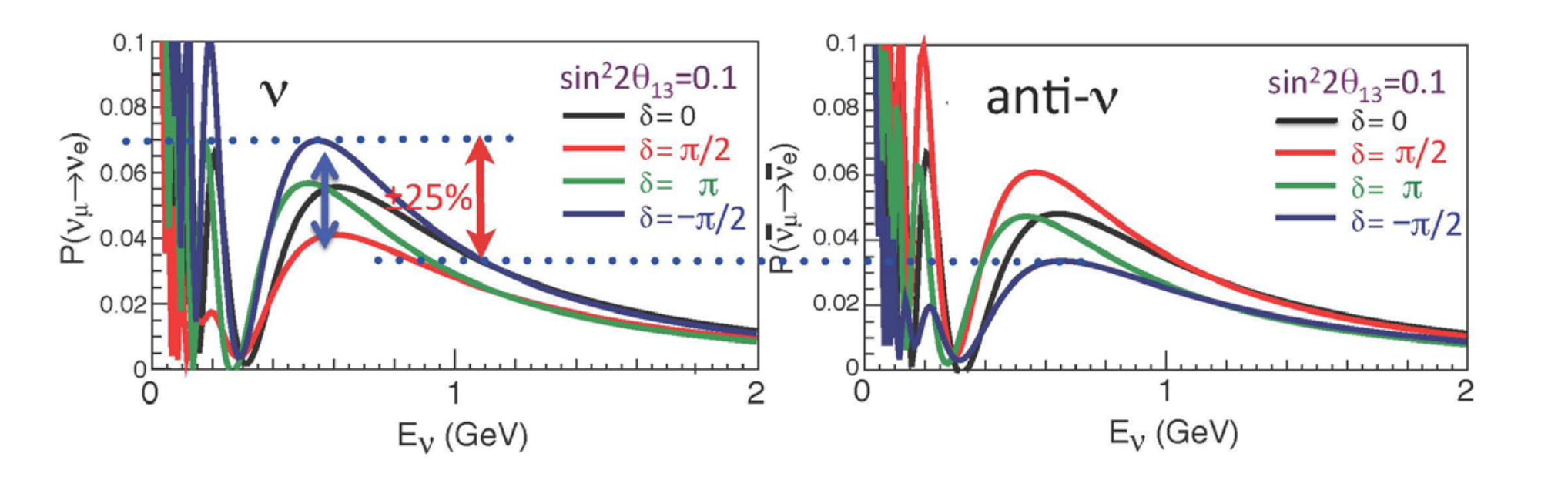}
\caption{\label{fig:difference}
Oscillation probabilities for $\nu_{\mu} \rightarrow \nu_{e}$ (left) and for
${\bar \nu_{\mu}} \rightarrow {\bar \nu_{e}}$ (right)  as functions of (anti-) neutrino energy.
The neutrino travel distance is 295~km.
The anti-neutrino oscillation probability is larger
(smaller) than the neutrino oscillation probability for positive (negative) $\delta_{\rm CP}$.
The maximum difference is approximately 25\%.
}
\end{figure}
\end{center}

The probability for ${\bar \nu_{\mu}} \rightarrow {\bar \nu_{e}}$ oscillation
can be written as follows;
$$
\begin{array}{lll}
P({\bar \nu_{\mu}} \rightarrow {\bar \nu_{e}}) &\approx & \sin^{2}\theta_{23} \sin^{2}2\theta_{13} 
      \sin^{2}\Bigl({{\Delta m^{2}_{32} L}\over{4 E_{\nu}}}\Bigr)
      \Bigl(1+{{4\sqrt{2}G_{F}n_{e}E}\over{\Delta m^{2}_{31}}}(1-2\sin^{2}\theta_{13})\Bigr)\nonumber\\
      &+& \sin 2\theta_{12} \sin 2\theta_{23} \sin 2\theta_{13} \cos\theta_{13} \sin\delta_{\rm CP}
      \sin^{2}\Bigl({{\Delta m^{2}_{32} L}\over{4 E_{\nu}}}\Bigr)
      \sin^{2}\Bigl({{\Delta m^{2}_{21} L}\over{4 E_{\nu}}}\Bigr).\\
\end{array}
$$
\noindent
It is slightly different from the probability for the $\nu_{\mu} \rightarrow \nu_{e}$ oscillation;
the sign of $\delta_{\rm CP}$ term is not minus but plus for ${\bar \nu_{\mu}} \rightarrow {\bar \nu_{e}}$ oscillation.
Because of this difference, the anti-neutrino oscillation probability is larger
(smaller) than the neutrino oscillation probability for positive (negative) $\delta_{\rm CP}$
by up to 25\% as shown in Figure~\ref{fig:difference}.
Accordingly, comparison of oscillation probabilities between neutrinos and anti-neutrinos
could determine the $\delta_{\rm CP}$ value.

\begin{center}
\begin{figure}[t!]
\includegraphics[width=37pc]{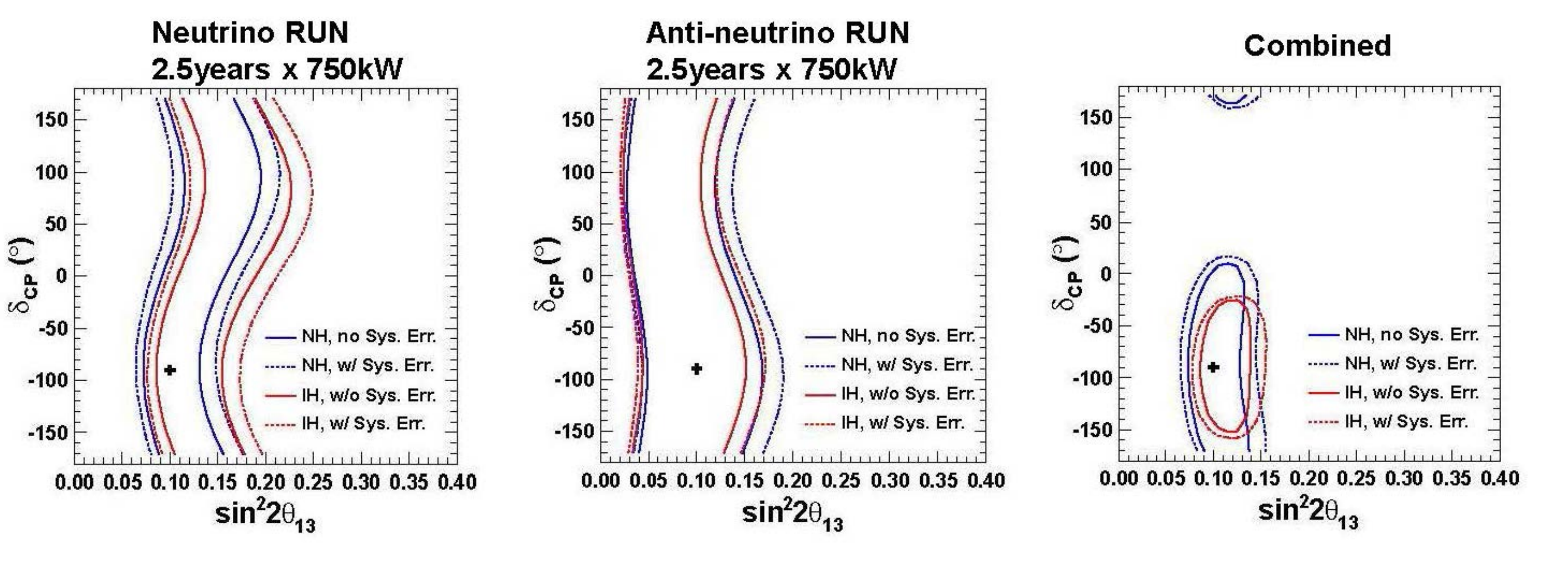}
\caption{\label{fig:potential}
Expected 90\% confidence level intervals in $\sin^{2}2\theta_{13}$ - $\delta_{\rm CP}$ planes.
The true parameter values are assumed to be
($\sin^{2}2\theta_{13}$,$\delta_{\rm CP}$)~=~(0.10, -$\pi/2$) with normal mass hierarchy,
as indicated by "+". 
The left (center) plot is the expectation 
with 2.5 years of 750kW neutrino (anti-neutrino) beam data.
The right plot is the result of combining neutrino and anti-neutrino beam data.
These plots were originally shown in Figure~3 of Ref~\cite{T2Kpotential}.  
}
\end{figure}
\end{center}

The T2K group is planning to assign 50\% of the beam time to the anti-neutrino beam run.
Future sensitivity studies based on this condition are provided in \cite{T2Kpotential}.
The point of the plan is well explained by Figure~\ref{fig:potential}.
The shapes of allowed regions in $\sin^{2}2\theta_{13}$ - $\delta_{\rm CP}$ planes are different
between neutrinos and anti-neutrinos.
By combining two allowed regions, constraints for the $\delta_{\rm CP}$ range
can be obtained in a single experiment without the use of reactor results.

As reported in section 3, the anti-neutrino beam started in June 2014, and $4.04\times 10^{20}$POT data
was accumulated until June 2015. In the following two sections, preliminary results
for anti-neutrino beam data are presented\cite{EPS2015}.
Although the final goal is a measurement of ${\bar \nu_{\mu}} \rightarrow {\bar \nu_{e}}$ oscillation
probability and a comparison with $\nu_{\mu} \rightarrow \nu_{e}$ probability, the first step is
confirmation of ${\bar \nu_{\mu}}$ disappearance and ${\bar \nu_{e}}$ appearance.

\vfill\eject

\section{Preliminary results of ${\bar \nu_{\mu}}$ disappearance}

\begin{center}
\begin{figure}[b!]
\includegraphics[width=38pc]{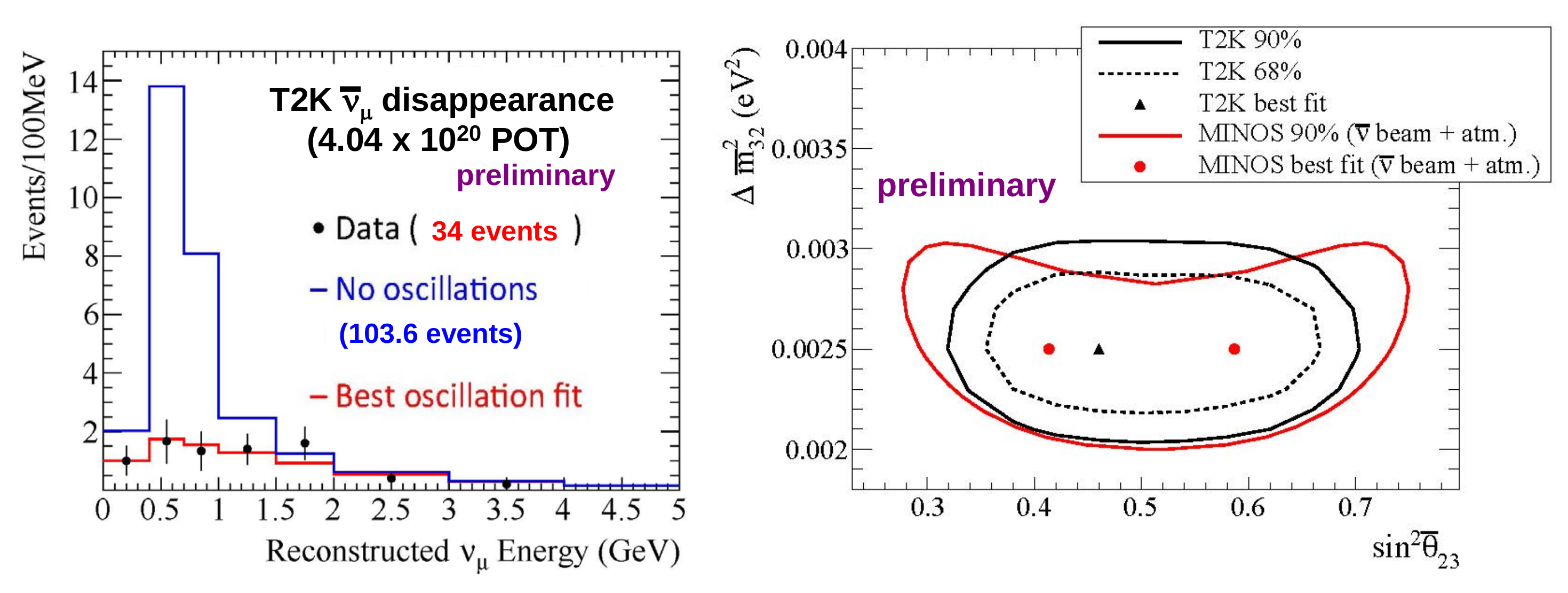}
\caption{\label{fig:numubar}
(left)~Distribution of reconstructed neutrino energy for 34 single ring
$\mu$-like events observed in
$4.04\times10^{20}$~POT anti-neutrino beam data. The expectations
for no oscillation and for best fit oscillation parameters are also shown.    
(right)~Constraints on-oscillation parameters $\Delta {\bar m^{2}_{32}}$ and $\sin^{2}{\bar \theta_{23}}$
obtained from ${\bar \nu_{\mu}}$ disappearance.
Constraints from the MINOS\cite{MINOSdisapp} experiment  are also shown.
}
\end{figure}
\end{center}

The disappearance of muon anti-neutrinos
is examined using $4.04\times 10^{20}$~POT anti-neutrino beam data.
The analysis procedure that includes the event selection and calculation of the
expectations by the Monte Carlo simulation are the same as those for neutrino analysis,
and are not repeated here. 
In this case 103.6 events are expected for no oscillation and 34 events are observed.
A reconstructed anti-neutrino energy spectrum is shown in Figure~\ref{fig:numubar}~(left).
The reduction in event numbers as well as the distortion of the energy spectrum is obvious.

The best-fit oscillation parameters were calculated as follows:
$$
\begin{array}{ll}
\Delta {\bar m^{2}_{32}} &= (2.50^{+0.3}_{-0.2}) \times 10^{-3} {\rm eV^{2}}\\
\sin^{2}{\bar \theta_{23}} &= 0.46^{+0.14}_{-0.06}\\
\end{array}
$$
\noindent
for normal mass hierarchy. No difference in best-fit parameters can be found for
inverted mass hierarchy.
No significant differences can be found between this section's results and those from
the disappearance of muon neutrinos reported in section 6. 

The constraints on oscillation parameters in $\sin^{2}{\bar \theta_{23}}$ - $\Delta {\bar m^{2}_{32}}$ plane
are calculated.
The results for normal mass hierarchy are shown in Figure~\ref{fig:numubar}~(right)
with results from the MINOS\cite{MINOSdisapp} experiment.
The T2K results are consistent with MINOS.
Again, more stringent constraints for $\sin^{2}{\bar \theta_{23}}$ are obtained by T2K.
\section{Preliminary results of ${\bar \nu_{e}}$ appearance}

The selection process of ${\bar \nu_{e}}$ candidates in SK is exactly
the same as that for neutrino beam data.
After all selections, three events remain as possible candidates of the
$\bar {\nu_{e}}$ appearance signal.

The absence of ${\bar \nu_{\mu}} \rightarrow {\bar \nu_{e}}$ oscillation is assumed,
and the expected numbers of background events are calculated.
The background events include 
$\nue$ appearance from $\nu_{\mu} \rightarrow \nu_{e}$ oscillation, misidentified $\nu_{\mu}$ (or  ${\bar \nu_{\mu}}$),
and original $\nu_{e}$(or $\bar \nu_{e}$) from the decay of muons in the T2K beam line.
In these calculations, the existence of $\nu_{\mu} \rightarrow \nu_{e}$ oscillation with
oscillation parameters by PDG2014\cite{PDG2014} is assumed, although this assumption
contradicts the absence of  ${\bar \nu_{\mu}} \rightarrow {\bar \nu_{e}}$
oscillation within the MNS framework.
The number of background events varies from 1.51 to 1.77, depending on mass hierarchy and $\delta_{\rm CP}$.
Obviously, the observation of three candidates is not significant evidence of ${\bar \nu_{e}}$ appearance.

The best-fit oscillation parameters obtained from neutrino beam analysis are assumed, and
the number of expected ${\bar \nu_{e}}$ candidates are calculated for some $\delta_{\rm CP}$ with normal/inverted
mass hierarchy. The results are 3.73$\sim$5.45 as summarized in Table~\ref{tab:antinueapp}.
All of these numbers are not excluded by the T2K observation of the three events.
Because of inadequate statistics, present T2K data cannot be used to examine any assumptions related to the oscillation
parameters.

\begin{table}[t!] 
\caption{
Number of expected ${\bar \nu_{e}}$ candidates for 4.04$\times$10$^{20}$POT T2K
anti-neutrino beam. The results from T2K data are also listed.
}
\smallskip
\begin{center}
\begin{tabular}{l|cc}
\hline
\hline
                            &~~~~~~~~~$\delta_{\rm CP}$~~~~~~& Number of  ${\bar \nu_{e}}$ candidates \\
\hline
Normal mass hierarchy &~~~~~~~~~$-\pi$/2~~~~~~& 3.73\\ 
                        &~~~~~~~~~0~~~~~~ & 4.32\\ 
                        &~~~~~~~~~$\pi$/2~~~~~~& 4.85\\ 
\hline
Inverted mass hierarchy &~~~~~~~~~$-\pi$/2~~~~~~&4.18\\ 
                        &~~~~~~~~~0~~~~~~& 4.85\\ 
                        &~~~~~~~~~$\pi$/2~~~~~~& 5.45\\ 
\hline
Data                 &  &    3 \\
\hline
\hline
\end{tabular}
\end{center}
\label{tab:antinueapp}
\end{table}

\section{Summary}
The T2K long-baseline neutrino-oscillation
experiment accumulated 
$6.57\times 10^{20}$ POT neutrino beam data
and 
$4.04\times 10^{20}$ POT anti-neutrino beam data
until June 2015.

In the $\nu_{\mu}$ disappearance analysis, 120 muon neutrino candidates are observed
where the expectation for no oscillation is $446\pm 23$.
The disappearance of $\nu_{\mu}$ candidate events and the distortion of the neutrino
energy spectrum are found. The best oscillation parameters were
$$
\begin{array}{ll}
\Delta m^{2}_{32} &= (2.51\pm0.10)\times 10^{-3} {\rm eV^{2}}\\
\sin^{2}\theta_{23} &= 0.514^{+0.055}_{-0.056}\\
\end{array}
$$
\noindent
for normal mass hierarchy.
Constraints in $\sin^{2}\theta_{23}$ - $\Delta m^{2}_{32}$ plane are consistent
with SK and MINOS.
The most stringent constraints are obtained for $\sin^{2}\theta_{23}$ by T2K.

In the $\nu_{e}$ appearance analysis,  28 electron neutrino candidates are observed
where $4.9\pm 0.6$ events are expected if no oscillation is assumed.
The significance of the signal is 7.3 times the standard deviation, and it is certainly discovery
of $\nu_{e}$ appearance.
Results from reactors are combined, and the parameter range

$$
\begin{array}{llll}
0.146\pi & < \delta_{\rm CP} < 0.825\pi & \rm{~~~~~~for~normal~mass~hierarchy~(NH)} &~~~~{\rm and}\\
&&&\\
-0.080\pi & < \delta_{\rm CP} < 1.091\pi & \rm{~~~~~~for~inverted~mass~hierarchy~(IH)} &\\
\end{array}
$$
\noindent
are excluded with a 90\% confidence level.
These results are considered to be the first hints toward
$\delta_{\rm CP} \sim -\pi/2$
and normal mass hierarchy.

In the ${\bar \nu_{\mu}}$ disappearance analysis, 34 ${\bar \nu_{\mu}}$ candidates are found
where 103.6 events are expected for no oscillation.
Distortion of the energy spectrum is also found, and the disappearance of ${\bar \nu_{\mu}}$ is obvious.
The constraints on oscillation parameters, $\Delta {\bar m^{2}_{32}}$ and $\sin^{2}{\bar \theta_{23}}$,
are consistent with the results from the MINOS experiment.
Any differences from parameters for neutrinos,  $\Delta m^{2}_{32}$ and $\sin^{2}\theta_{23}$,
are identified.

In the ${\bar \nu_{e}}$ appearance analysis, three ${\bar \nu_{e}}$ candidates are found, where
the numbers of background events are 1.51$\sim$ 1.77, depending on mass hierarchy and $\delta_{\rm CP}$.
Expected ${\bar \nu_{e}}$ candidates are calculated for some $\delta_{\rm CP}$ with normal/inverted
mass hierarchy, and are 3.73$\sim$5.45.
Present ${\bar \nu_{e}}$ appearance analysis cannot provide any constraints for
the oscillation parameters, because of inadequate statistics.

\vfill\eject

\end{document}